\input harvmac

\input psbox

\def\inbar{\,\vrule height1.5ex width.4pt depth0pt}
\def\IB{\relax{\rm I\kern-.18em B}}
\def\IC{\relax\hbox{$\inbar\kern-.3em{\rm C}$}}
\def\ID{\relax{\rm I\kern-.18em D}}
\def\IE{\relax{\rm I\kern-.18em E}}
\def\IF{\relax{\rm I\kern-.18em F}}
\def\IG{\relax\hbox{$\inbar\kern-.3em{\rm G}$}}
\def\IH{\relax{\rm I\kern-.18em H}}
\def\II{\relax{\rm I\kern-.18em I}}
\def\IK{\relax{\rm I\kern-.18em K}}
\def\IL{\relax{\rm I\kern-.18em L}}
\def\IM{\relax{\rm I\kern-.18em M}}
\def\IN{\relax{\rm I\kern-.18em N}}
\def\IO{\relax\hbox{$\inbar\kern-.3em{\rm O}$}}
\def\IP{\relax{\rm I\kern-.18em P}}
\def\IQ{\relax\hbox{$\inbar\kern-.3em{\rm Q}$}}
\def\IR{\relax{\rm I\kern-.18em R}}
\font\cmss=cmss10 \font\cmsss=cmss10 at 7pt
\def\IZ{\relax\ifmmode\mathchoice
{\hbox{\cmss Z\kern-.4em Z}}{\hbox{\cmss Z\kern-.4em Z}}
{\lower.9pt\hbox{\cmsss Z\kern-.4em Z}}
{\lower1.2pt\hbox{\cmsss Z\kern-.4em Z}}\else{\cmss Z\kern-.4em Z}\fi}
\def\IGa{\relax\hbox{${\rm I}\kern-.18em\Gamma$}}
\def\IPi{\relax\hbox{${\rm I}\kern-.18em\Pi$}}
\def\ITh{\relax\hbox{$\inbar\kern-.3em\Theta$}}
\def\IOm{\relax\hbox{$\inbar\kern-3.00pt\Omega$}}

\font\cmss=cmss10

\Title{\vbox{\baselineskip12pt\hbox{UICHEP-TH/96-02}}}
{\vbox{\centerline{Topology Change and $\theta$-Vacua in 2D 
Yang-Mills Theory}}}

\centerline{Tom D. Imbo$^{*,}$\footnote{$^1$}{imbo@uic.edu} 
and Paulo Teotonio-Sobrinho$^{\dagger ,}$\footnote{$^2$}
{teotonio@fma.if.usp.br}}

\bigskip\centerline{$^*$Department of Physics}
\centerline{University of Illinois at Chicago}
\centerline{845 W. Taylor St.}
\centerline{Chicago, IL \ 60607-7059, USA}
\bigskip\centerline{$^{\dagger}$Instituto de Fisica -- DFMA}
\centerline{Universidade de Sao Paulo}
\centerline{Caixa Postal 66318, 05315-970}
\centerline{Sao Paulo, SP, Brazil}
\vskip 0.7in

We discuss the existence of $\theta$-vacua in pure Yang-Mills theory in
two space-time dimensions. More precisely, a procedure is given which
allows one to classify the distinct quantum theories possessing the same
classical limit for an arbitrary connected gauge group $G$ and compact
space-time manifold $M$ (possibly with boundary) possessing a special 
basepoint. For any such $G$ and $M$ it is shown that the above 
quantizations are in one-to-one correspondence with the irreducible 
unitary representations (IUR's) of $\pi_1(G)$ if $M$ is orientable, and 
with the IUR's of $\pi_1(G)/2\pi_1(G)$ if $M$ is nonorientable. 

\Date{}

Theories of interacting gauge and matter fields in two-dimensional
space-time bear enough resemblance to their four-dimensional counterparts
to make them interesting theoretical laboratories within which to
investigate various qualitative effects, as well as test certain
approximation schemes. However, in contrast to the situation in dimension
three or more, there are no propagating gauge bosons in two dimensions. 
One may therefore expect that {\it pure} gauge theories in two dimensions 
are trivial. But if the two-dimensional space-time $M$ has a nontrivial 
topology, then this is not quite true. For example, let's consider the 
case where $M$ has the topology of a cylinder, the ``space manifold'' 
being simply a circle. Then the nontrivial fixed-time degrees of freedom 
are the path-ordered exponentials (or {\it holonomies}) of the gauge fields 
along a contour which winds once around the spatial circle, and the theory 
can be shown to be equivalent to the quantum mechanics of a free particle 
moving on the gauge group $G$ \ref\raj{S.~G.~Rajeev, Phys. Lett. {\bf B212} 
(1988) 203\semi J.~E.~Hetrick and Y.~Hosotani, Phys. Lett. {\bf B230} (1989) 
88\semi S.~Guruswamy and S.~G.~Rajeev, Mod. Phys. Lett. {\bf A7} (1992) 
3783\semi E.~Langmann and G.~W.~Semenoff, Phys. Lett. {\bf B296} (1992) 117; 
{\bf B303} (1993) 303\semi J.~E.~Hetrick, Int. J. Mod. Phys. {\bf A9} (1994)
3153.}\ref\chan{L.~Chandar and E.~Ercolessi, Nucl. Phys. {\bf B426} (1994) 
94.}. (There are some subtleties here concerning boundary conditions which 
we will address later.) The radius $R$ of the circle and the gauge coupling 
constant determine the parameters of this equivalent system. In particular, 
the volume of ``space'' (that is, $G$) in the quantum mechanical model 
decreases as $R$ increases. In the limit $R\to\infty$ this volume tends to 
zero, causing all energy level differences to grow without bound and leaving 
us with a trivial theory as expected. Other topologically interesting 
space-times do not admit such a nice canonical decomposition into 
(space-manifold) $\times$ (time). Consequently, pure gauge theories on these 
manifolds do not have an interpretation as ordinary quantum mechanical 
models as above, although they are still ``less than field theories''. 
Indeed, they are all ``almost topological'' in the sense that the only 
property of the metric on $M$ that the theory depends on is the area. 
Recently, the exact partition functions of these models have been obtained
for an arbitrary space-time and gauge group \ref\part{A.~A.~Migdal, Sov. Phys. 
JETP {\bf 42} (1975) 413\semi B.~Ye.~Rusakov, Mod. Phys. Lett. {\bf A5} (1990) 
693\semi E.~Witten, Comm. Math. Phys. {\bf 141} (1991) 153; J. Geom. Phys. 
{\bf 9} (1992) 303\semi M.~Blau and G.~Thompson, Int. J. Mod. Phys. {\bf A7} 
(1992) 3781.}. These were then used to show that in the large-N limit, any 
pure U(N), SU(N), SO(N), or Sp(N) gauge theory on a closed two-dimensional 
space-time is equivalent to a closed string theory \ref\str{D.~Gross, Nucl. 
Phys. {\bf B400} (1993) 161\semi D.~Gross and W.~Taylor, Nucl. Phys. 
{\bf B400} (1993) 181; Nucl. Phys. {\bf B403} (1993) 395\semi J.~Minahan, 
Phys. Rev. {\bf D47} (1993) 3430\semi S.~Naculich, H.~Riggs and H.~Schnitzer, 
Mod. Phys. Lett. {\bf A8} (1993) 2223; Phys. Lett. {\bf B319} (1993) 466\semi 
S.~Ramgoolam, Nucl. Phys. {\bf B418} (1994) 30.}. This holds for any value of 
the gauge coupling constant, except when space-time is the sphere $S^2$ or 
the projective plane $P^2$ in which case there is a third-order phase 
transition (from a trivial to a stringy phase) as one moves from weak to 
strong coupling \ref\phase{M.~Douglas and V.~Kazakov, Phys. Lett. {\bf B319} 
(1993) 219\semi J.~Minahan and A.~Polychronakos, Nucl. Phys. {\bf B422} (1994) 
172\semi D.~Gross and A.~Matytsin, Nucl. Phys. {\bf B429} (1994) 50; 
{\bf B437} (1994) 541\semi M.~Crescimanno and H.~Schnitzer, Int. J. Mod. 
Phys. {\bf A11} (1996) 1733.}. These results lend support to the long-held 
belief that there is a stringy formulation of four-dimensional nonabelian 
gauge theories.

Another interesting feature that gauge theories in two and four dimensions
share is the existence of $\theta$-{\it vacua}. That is, there exist in
general numerous quantizations of the theory which possess the same
classical limit. The name ``$\theta$-vacua'' derives from QCD in ordinary
four-dimensional Minkowski space, where the above quantum theories are
labelled by an angle $\theta$ and can be implemented by adding the
well-known Pontryagin density (with coefficient $\theta$) to the
Lagrangian. A single angle also labels the quantizations of a $U(1)$
gauge theory in two-dimensional Minkowski space (the {\it Schwinger
model}) \ref\csm{S.~Coleman, R.~Jackiw and L.~Susskind, Ann. Phys. {\bf
93} (1975) 267\semi S.~Coleman, Ann. Phys. {\bf 101} (1976) 239.}, where
the topological term in the Lagrangian is now ${\theta\over 4\pi}
\epsilon^{\mu\nu}F_{\mu\nu}$. For {\it nonabelian} gauge theories 
in two-dimensional Minkowski space with a semi-simple gauge group, there is 
no longer a continuous parameter labelling the inequivalent quantum theories. 
There are now only a discrete number of choices available, parametrized by 
the distinct representations of the gauge group $G$. Moreover, there is no 
local term that can be added to the Lagrangian that can implement any of 
these choices. Instead, a multiplicative term in the quantum mechanical 
path integral must be inserted. More precisely, in order to obtain the 
quantization associated to a given representation $D$ of $G$, one must 
multiply the path-integrand by the Wilson loop of the gauge field around 
the circle at space-time infinity, where the trace in the Wilson loop is 
evaluated in the representation $D$ \ref\wit{E.~Witten, Nuov. Cim. 
{\bf 51 A} (1979) 325.}. Not every distinct choice of $D$ will yield a 
distinct quantum theory. The number of inequivalent theories depends on 
both the gauge group $G$ and the matter content of the model. (In 
two-dimensional Minkowski space, which has the topology of the plane, 
some matter fields are necessary in order to obtain a nontrivial theory
as noted earlier.) For example, if $G=SU(n)$ and the model contains 
quarks in the fundamental representation, then all choices of $D$ yield 
the same theory. If the quarks are instead in the adjoint representation, 
then there are $n$ distinct quantizations (or vacua) \wit . In recent 
years, there have been numerous investigations into the properties of 
these multiple vacua in 2D gauge theories (see \ref\web{L.~D.~Paniak, 
G.~W.~Semenoff and A.~R.~Zhitnitsky, hep-th/9606194\semi A.~V.~Smilga, 
hep-th/9607007.}, and references therein).

All of these Minkowski space results, including the standard results in
four dimensions, can be understood by considering the topology of the
gauge theory configuration space (see \ref\bal{A.~P.~Balachandran, G.~Marmo,
B.-S.~Skagerstam and A.~Stern, {\it Classical Topology and Quantum States}
(World Scientific, Singapore, 1991).}, and references therein). In a pure 
gauge theory in ($d$+1)-dimensions with connected gauge group $G$, the 
classical configuration space consists of the set ${\cal A}$ of all 
fixed-time, finite-energy gauge fields (in temporal gauge). However, we must 
identify any two elements of ${\cal A}$ that differ by a static local gauge 
transformation which is trivial at spatial infinity. More precisely, the set 
of all such gauge transformations form an infinite-dimensional Lie group under 
pointwise multiplication. This {\it group of gauge transformations}, which we 
denote by ${\cal G}$, acts on $\cal A$ in the usual way. The actual 
configuration space of the theory is the orbit space $Q={\cal A}/{\cal G}$. 
The restriction that gauge transformations in ${\cal G}$ are asymptotically 
trivial assures us that gauge fields differing by a nontrivial {\it global} 
gauge transformation are treated as distinct. It also makes the action of 
${\cal G}$ on ${\cal A}$ free (that is, no gauge field is fixed under any 
nontrivial element of ${\cal G}$), 
so that $Q$ is a smooth manifold. This boundary 
condition in effect compactifies the space manifold $\IR^d$ to the $d$-sphere 
$S^d$, which comes equipped with a special {\it basepoint} --- the ``point at 
infinity''. The wave functionals composing the Hilbert space of any given 
quantization of the system can be viewed as sections of a fixed complex vector 
bundle over $Q$. The quantum theories associated with the {\it flat} bundles 
all have the same classical limit, and are in one-to-one correspondence with 
the distinct (up to overall conjugation) unitary representations of the 
fundamental group $\pi_1(Q)$. It suffices to consider only the irreducible 
unitary representations (IUR's) of $\pi_1(Q)$ since all other quantizations 
can be easily obtained from these ``irreducible'' ones. In what follows, 
by a {\it quantization} we will mean a quantum theory constructed using 
one of these irreducible flat vector bundles over $Q$.

It is straightforward to show that ${\cal A}$ is not, in general, a 
path-connected space. The path-components (or {\it soliton sectors}) 
${\cal A}_{\alpha}$ of $\cal A$ are labelled by the elements $\alpha\in
\pi_{d-1}(G)$ --- that is, $\pi_0({\cal A})=\pi_{d-1}(G)$. These different 
sectors correspond to topologically distinct behaviors of the gauge fields 
at spatial infinity. Since the elements of $\cal G$ are trivial at spatial 
infinity, modding out by $\cal G$ does not affect this classification. In 
other words, we also have $\pi_0(Q)=\pi_{d-1}(G)$. It can further be shown 
that each path-component of $\cal A$ is contractible --- that is, for each 
$\alpha\in\pi_{d-1}(G)$ we have $\pi_m({\cal A}_{\alpha})=\{ e\}$ for all 
$m\geq 1$. This fact, combined with the freeness of the ${\cal G}$ action 
and some standard results from algebraic topology, suffices to show that 
$\pi_m(Q_{\alpha})=\pi_{m-1}({\cal G})=\pi_{m+d-1}(G)$ for all $\alpha\in
\pi_{d-1}(G)$ and $m\geq 1$. In particular, we have that $\pi_1(Q_{\alpha})=
\pi_d(G)$ for all $\alpha$. Thus, the inequivalent quantizations of the gauge 
theory (in any soliton sector) are labelled by the IUR's of $\pi_d(G)$. When 
we add ordinary matter fields to the theory, the configuration space will 
clearly change. However, it can be shown that the homotopy groups of $Q$, in 
particular the fundamental group, are unaffected. The matter content of the 
theory {\it may} affect, however, the choice of gauge group $G$ which should 
be used in the above analysis. For example, in a pure gauge theory the fields 
(which lie in the Lie algebra of $SU(n)$) cannot detect the center of $SU(n)$, 
which is isomorphic to the cyclic group $\IZ_n$. Therefore, the natural 
choice for $G$ is the quotient group $SU(n)/\IZ_n$. This result doesn't 
change if we add adjoint matter to the theory since these fields are also 
unaffected by central elements. But the {\it fundamental representation} of 
$SU(n)$ is faithful, so that the addition of fields transforming according to 
this representation leads us to choose $G=SU(n)$. Other choices of matter 
representations will yield, in general, different quotient groups of $SU(n)$. 

Since $\pi_3(G)$ is isomorphic to the additive group of the integers $\IZ$
when $G$ is simple and nonabelian, we see that in three spatial dimensions
there are a continuum of inequivalent quantizations of the corresponding
gauge theories. More precisely, the IUR's of $\IZ$ are all one-dimensional
and are labelled by an angle $\theta$ --- the IUR corresponding to a given
$\theta$ is given by $n\to e^{{\rm i}n\theta}$ for all $n\in\IZ$. These
yield the well-known $\theta$-vacua discussed previously. By contrast,
since $\pi_3(U(1))$ is trivial there is no such phenomenon in
(3+1)-dimensional QED. Also, the triviality of $\pi_2(G)$ for all Lie
groups implies that there are no $\theta$-vacua in (2+1)-dimensions.
However, since $\pi_1(U(1))=\IZ$, we see the existence of $\theta$-vacua
in (1+1)-dimensional $QED$ as noted above. Simply-connected gauge groups 
($\pi_1(G)=\{e\}$) such as $SU(n)$ do not lead to any such quantization 
ambiguities in 2D, while many other semi-simple gauge groups (such as $SO(n)$ 
or $SU(n)/\IZ_n$) have a finite fundamental group ($\pi_1(SO(n))=\IZ_2$ and
$\pi_1(SU(n)/\IZ_n)=\IZ_n$). These latter groups possess the discrete set
of quantizations alluded to earlier.

The above results also hold in the case where the space manifold is
not $\IR^d$, but the solid $d$-dimensional ball $B^d$ of some finite radius
$R$. This is true so long as the boundary conditions imposed on the gauge 
fields and gauge transformations on the ($d-1$)-dimensional spherical boundary 
of $B^d$ are similar to the corresponding conditions imposed at spatial 
infinity in the $\IR^d$ case --- in effect compactifying $B^d$ to $S^d$ with 
a special basepoint.\foot{Even though the configuration spaces have the 
same topology in these two cases, the {\it dynamics} of the two theories can 
be quite different. They must, however, coincide as $R\to\infty$.} For $d=1$, 
space is a simply a line segment which is compactified to a circle. The 
basepoint $y_0$ of this circle represents the two endpoints of the line 
segment which are equated by the boundary conditions. Note that the holonomy 
of any static gauge field $A$ around this spatial circle (based at $y_0$) is 
invariant under the elements of ${\cal G}$, which are required to be trivial 
at $y_0$. Moreover, specifying this holonomy is equivalent to specifying the 
gauge equivalence class of $A$ in the configuration space $Q$. Since these 
holonomies can be any element of the gauge group $G$, we see that $Q$ is 
homeomorphic to the group manifold of $G$. (In particular, we recover 
$\pi_0(Q)=\pi_0(G)$ and $\pi_1(Q)=\pi_1(G)$.) By contrast, $Q$ is 
infinite-dimensional for all $d>1$ since here we have a true field theory --- 
that is, there {\it are} propagating gauge bosons.

There is a subtlety here concerning boundary conditions. Namely, the way in 
which we obtained a cylindrical space-time above was to compactify space from 
a line segment to a circle. But we may also wish to consider the case where 
space is simply a circle at the outset, and not just as a result of boundary 
conditions. The difference in these two situations lies in the existence of 
a basepoint. Saying that space is {\it a priori} 
compact means that there is no 
special point on the circle chosen by the physics --- all points are on the 
same footing. As a result, there is no natural place to require the elements 
of ${\cal G}$ to be trivial. If we simply choose a random point at
which to do so, this will 
violate our initial assumption about the nature of the space manifold. In 
other words, we will be changing the physics. But if we do not have such a 
point, then we forced to include {\it all} static gauge transformations in 
${\cal G}$, even global ones. Since holonomies conjugate under global gauge 
transformations, the configuration space $Q=G$ obtained in the ``pointed'' 
case above is then replaced by the orbit space $G/G$, where $G$ acts on itself 
by conjugation. That is, $Q$ is now the space of {\it conjugacy classes} in 
$G$. (For instance, the space of conjugacy classes of $G=SO(3)$ is 
homeomorphic to a line segment.) Understanding the inequivalent quantizations 
of these theories is a bit more involved \chan . However, there remains a 
subclass of quantizations which are labelled by the unitary representations of 
$\pi_1(G)$. In what follows, we will always assume a special basepoint and 
avoid these complications.

The above canonical methods generalize to any space-time which is homeomorphic 
to $\Sigma\times I$, where $\Sigma$ is a space-like manifold of dimension 
$d\geq 1$ and $I=[0,1]$ is the unit interval on the real line. One simply 
replaces $\cal A$ by the space of all static gauge fields on $\Sigma$, and 
${\cal G}$ by the associated group of static
gauge transformations (both 
satisfying restrictions appropriate to the topology of $\Sigma$). The 
configuration space $Q(\Sigma )$ is again the corresponding orbit space. The 
inequivalent quantizations of the theory are still labelled by the IUR's of 
$\pi_1(Q(\Sigma ))$. However, the computation of $\pi_1(Q(\Sigma ))$ is 
somewhat more involved for a general $\Sigma$. In particular, 
the result is no longer simply
$\pi_d(G)$ in general \ref\ti{C.~J.~Isham, Phys. Lett. {\bf B106} 
(1981) 188\semi E.~C.~G.~Sudarshan, T.~D.~Imbo and T.~R.~Govindarajan, Phys. 
Lett. {\bf B213} (1988) 471\semi T.~D.~Imbo, Ph.~D. Dissertation, University 
of Texas at Austin, 1988.}.

We now turn to the main purpose of this paper, which is to determine the 
inequivalent quantizations of a pure Yang-Mills theory on a {\it general} 
two-dimensional compact space-time $M$ (possibly with boundary). We are 
particularly interested in cases where there is no nonsingular global 
slicing of $M$ into space-like manifolds, since the canonical techniques 
described earlier do not apply here. For example, $M$ can be the two-sphere 
$S^2$ (or any closed surface other than the torus), or have the ``pair of 
pants'' topology shown in Fig.1. Such a situation is often said to 
involve {\it topology change} since any attempt to foliate $M$ into 
space-like pieces will necessarily have special slices where space is not a 
manifold (there will be singular regions in space). Immediately before and 
immediately after these special time values the spatial slices {\it will} be 
manifolds in general, but their topologies may be quite different. In the 
example of Fig.1, the spatial topology can be viewed as changing from that 
of a circle to that of a disjoint union of two circles through a singular 
space having the topology of a figure eight. Since fixed-time degrees of 
freedom are no longer well-defined globally, we must find a covariant analog 
of the canonical configuration space $Q(\Sigma )$ above. We will use the 
{\it space of histories} which differs from $Q(\Sigma )$ essentially by 
replacing $\Sigma$ by the full space-time manifold $M$. More precisely, 
${\cal A}$ is replaced by the set ${\cal A}^H$ of all space-time gauge fields, 
and ${\cal G}$ is replaced by the group ${\cal G}^H$ of all space-time 
gauge transformations. Again, the elements of both ${\cal A}^H$ and 
${\cal G}^H$ are assumed to satisfy restrictions appropriate to the topology 
of $M$. The space of histories is then the orbit space $Q^H(M)={\cal A}^H/
{\cal G}^H$. However, since ``time'' has already been 
folded into the definition 
of $Q^H(M)$, the analog of fundamental group $\pi_1(Q(\Sigma ))$ is the 
set $\pi_0(Q^H(M))$ which labels the path-components 
(or {\it instanton sectors}) 
of $Q^H(M)$. A quantization of the theory is then associated with some
map $\chi$ from $\pi_0(Q^H(M))$ to a unitary matrix group. Since 
$\pi_0(Q^H(M))$ is not endowed with a natural 
group structure in general, it no 
longer makes sense to require this map to be a homomorphism. But there
{\it are}  
certain restrictions on $\chi$. First, if $M$ is globally of the form 
$\Sigma\times I$ with $\Sigma$ a space-like manifold (that is, there is no 
topology change), then it is easy to show that $\pi_0(Q^H(\Sigma\times I))$ is 
in one-to-one correspondence with $\pi_1(Q(\Sigma ))$. In order to recover the 
canonical results, we must then consider only those $\chi$'s which yield 
representations of $\pi_1(Q(\Sigma ))$ under any one such correspondence. All 
other $\chi$'s are to be discarded.\foot{Changing the one-to-one
correspondence  
${\eta:\pi_0(Q^H(\Sigma\times I))\to\pi_1(Q(\Sigma ))}$ which is used in this 
procedure will, in general, change the set of allowed $\chi 's$. However, by 
construction, this set remains in one-to-one correspondence with the 
representations of $\pi_1(Q(\Sigma ))$ for any choice of $\eta$.} More
generally,  
if there is some region of $M$ of the form $\Sigma\times I$, then we
can always  
find a continuous map ${\phi :Q^H(\Sigma\times I)\to Q^H(M)}$ which uses some 
fixed prescription to extend any history on $\Sigma\times I$ to all of $M$. In 
turn, this will induce a map ${\phi_*:\pi_0(Q^H(\Sigma\times I))\to 
\pi_0(Q^H(M))}$ which relates local information about $\pi_0(Q^H)$ from such a 
canonical region to the corresponding global information obtained
utilizing the  
full structure of $M$. Using the previous result, we may rewrite this as 
${\phi_*:\pi_1(Q(\Sigma))\to\pi_0(Q^H(M))}$. Any quantization of the full 
theory must yield an allowed quantization in each such local
region. Therefore,  
we should only consider $\chi$'s such that for each canonical region of $M$ as 
above the composite map $\chi\circ\phi_*$ is a homomorphism (for some fixed 
prescription for constructing $\phi$ and some fixed one-to-one correspondence 
between $\pi_0(Q^H(\Sigma\times I))$ and $\pi_1(Q(\Sigma ))$). In other words, 
these $\chi$'s label the inequivalent quantizations. We can still identify the 
``irreducible'' quantizations from which all others can be built as those for 
which the matrices $\chi (\pi_0(Q^H(M)))$ constitute an irreducible set. A few 
examples should help clarify these issues. 

\centinsert{
   \pscaption{
        \psannotate{\psboxto(0cm;7cm){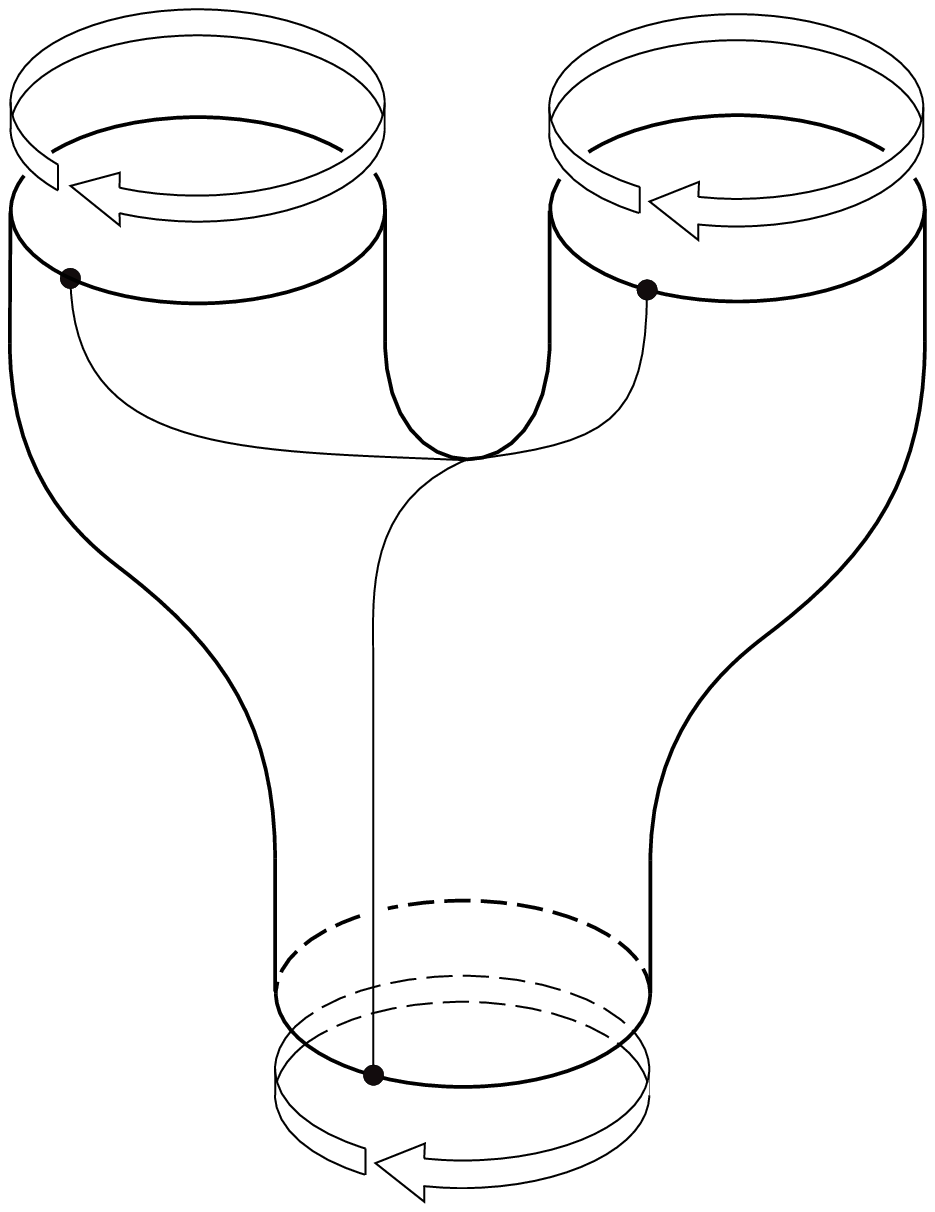}}
        {
         \at(12.5\pscm;0\pscm){$l_3$}
         \at(9.5\pscm;14\pscm){$l_1$}
         \at(15\pscm;14\pscm){$l_2$}
         \at(11.7\pscm;2.7\pscm){$y_3$}
         \at(14\pscm;9.5\pscm){$y_2$}
         \at(6\pscm;10\pscm){$y_1$}
        }}
   {Fig. 1: The ``pair of pants'' space-time $P$. The loops $l_1,l_2$, and 
$l_3$ (based at the points $y_1$, $y_2$ and $y_3$) are used to define the 
boundary holonomies $g_1,g_2$, and $g_3$. The ``history'' of the basepoints 
is also indicated. It consists of a network having a ``Y'' shape. }}

First, let's return to the case of a pure gauge theory, with connected gauge 
group $G$, defined on a space-time cylinder $C=S^1\times I$ where the 
spatial circle has a special basepoint. As noted earlier, the canonical 
configuration space $Q(S^1)$ is homeomorphic to the group manifold of
$G$. Even  
though this theory contains no topology change, let us repeat the
analysis using  
the space of histories $Q^H(C)$. By following the basepoint of the
spatial circle  
through time, we obtain a curve in space-time starting at one end of the 
cylinder and ending at the other. We parametrize this curve by $y_t$, $t\in 
[0,1]$. At each such time $t$ the holonomy of a gauge field configuration in 
${\cal A}^H$ around the spatial circle (based at $y_t$) yields an element 
$g(t)\in G$. Since we again wish to treat fields differing by a global gauge 
transformation as distinct, we require each gauge transformation in 
${\cal G}^H$ to be trivial at $y_t$ for all $t$.\foot{Actually, we only 
need to require this at any one time. However, our stronger assumption 
simplifies the analysis without changing the final result. More on
this later.}  
Each of the above holonomies $g(t)$ is then invariant under the elements of 
${\cal G}^H$, and when taken together contain the same information as does the 
${\cal G}^H$ equivalence class of the underlying space-time gauge
field. We must  
also fix the boundary holonomies on the two circular ends of $C$ at $t=0$ and 
$t=1$. We take these to be $g_1$ and $g_2$ respectively. Thus, $Q^H(C)$ can be 
viewed as the space of all maps $g:I\to G$ satisfying $g(0)=g_1$ and
$g(1)=g_2$.  
We can also view the theory as a (0+1)-dimensional ``nonlinear sigma model'' 
with target space $G$ --- or equivalently, the quantum mechanics of a single 
particle on $G$ as noted above. Since $Q^H(C)$ is nothing but the set of all 
continuous paths in $G$ between $g_1$ and $g_2$, we also see that
$\pi_0(Q^H(C))$ is in one-to-one correspondence with 
$\pi_1(G)$ (for any choice of $g_1$ and $g_2$). Recall that we  
have previously shown $\pi_1(Q(S^1))=\pi_1(G)$, so that we indeed obtain 
the same results in the new covariant approach as we did in the canonical 
picture described earlier. 

What happens if we now change the space-time manifold from the cylinder to
the manifold $P$ shown in Fig.1? Near each of the three ends
of $P$, space-time appears again to be simply a cylinder. Therefore,
restricting ourselves to any one of these regions we should find the same
quantization ambiguities labelled by the IUR's of $\pi_1(G)$ as above. The
question now becomes: Does anything interesting occur when we attempt to
paste these three regions together? That is, what labels the inequivalent
quantizations on the full space-time? If we follow the spatial basepoint
through ``time'' in this case, we no longer get a curve homeomorphic to
the unit interval. Instead we obtain something akin to the letter ``Y''.
In other words, the two distinct basepoints on the two upper legs of
space-time merge at the singular junction and become the basepoint of the
lower leg. We require the elements of ${\cal G}^H$ to be trivial along this 
entire ``basepoint network''. The elements of $Q^H(P)$ are then maps from 
this ``Y'' into $G$. Each such map, when evaluated at a point $y\in 
{\rm Y}$, represents the holonomy (based at $y$) of the gauge field around 
the spatial circle which contains $y$. There is, however, one exception to 
this. If $y$ is the junction point, then the holonomy is evaluated along 
the entire spatial figure-eight. Each element of $Q^H(P)$ is subject to 
boundary conditions which fix the values of the holonomy at all three ends. 
There is also one additional constraint --- an extra condition on the 
holonomies at the junction. In order to determine this condition, imagine 
starting with a loop going once around the spatial circle at each of the 
three ends of $P$. Now move each of these loops toward the singular time 
value. It is clear that the loop starting on the lower leg of space-time 
becomes the composite of the other two loops at this special time slice. 
This means that the lower leg holonomy must equal the product of the two 
upper leg holonomies in the limit as we approach the junction of the ``Y''. 
We may thus view $Q^H(P)$ as follows. First, consider a single copy of the 
gauge group $G$ marked with three special points $g_1$, $g_2$ and $g_3$. 
These special elements of $G$ represent the fixed holonomies of the gauge 
fields around the loops $\ell_1$, $\ell_2$ and $\ell_3$ at the ends of $P$ 
(see Fig.1). An element of $Q^H(P)$ then corresponds to a choice of three 
paths in $G$ emanating from $g_1$, $g_2$ and $g_3$ and ending at some 
points $h_1$, $h_2$ and $h_3$ respectively, subject to the constraint
$h_3=h_1h_2$. 

Now that we have a simple picture of $Q^H(P)$ at our disposal, we can
attempt to compute $\pi_0(Q^H(P))$ and use this to classify the
inequivalent quantizations of the system. Our claim is that the result is
identical to that obtained above for $Q^H(C)$ --- that is,
$\pi_0(Q^H(P))=\pi_1(G)$ independent of the choice of boundary holonomies 
on $P$, and the inequivalent quantizations are labelled by the IUR's of 
$\pi_1(G)$. In other words, the introduction of topology
change has not affected the number of distinct quantum theories. 
In order to prove this, we will need to relate  
the spaces $Q^H(P)$ and $Q^H(C)$. But first we must fix our boundary 
holonomies. On $P$ we stick to the generic choices $g_1$, $g_2$ and 
$g_3$ discussed previously. On $C$ we fix the gauge field holonomies
to be $g_3$ on the ``bottom'' end of the cylinder, and $g_1g_2$ on the top. 
(The reason for this funny choice will be made clear momentarily, but 
remember that $\pi_0(Q^H(C))=\pi_1(G)$ for any choice of boundary holonomies.) 
We will now define a continuous map $\phi :Q^H(C)\to Q^H(P)$. 
If $q_c\in Q^H(C)$, we let $\phi (q_c)$ be the history in $Q^H(P)$ which is
equal to $q_c$ on the lower leg of the ``Y'' (corresponding to the lower
cylindrical leg of $P$), and is constant and equal to $g_1$ (respectively, 
$g_2$) on the upper left (respectively, right) leg. Thus, $\phi (q_c)$ only 
changes with time in the period before the singularity. (Note that the 
constraint on the holonomies at the junction led to our choice of boundary 
conditions on $C$.) Clearly, $\phi$ is a one-to-one map. That is, $Q^H(C)$ is 
homeomorphic to the subspace $\phi (Q^H(C))$ of $Q^H(P)$. We will now show 
that each history in $Q^H(P)$ can be continuously deformed into this subspace. 
As noted earlier, any $q_p\in Q^H(P)$ may be viewed as a set of three paths in 
$G$ starting at $g_1$, $g_2$ and $g_3$ and ending at some points $h_1$, $h_2$ 
and $h_3$ respectively. These paths correspond to the holonomies on the three 
legs of $P$ and must satisfy the junction constraint $h_3=h_1h_2$. We will 
first deform the path starting at $g_1$ back along itself until it is simply 
remains at the point $g_1$. While doing this we will keep the path starting at 
$g_2$ fixed, so that in order to meet the junction constraint the path
starting  
at $g_3$ must simultaneously deform. When we are done, this path will end at 
$g_1h_2$. We now do the same deformation of the path starting at
$g_2$, keeping  
the (constant) path at $g_1$ fixed, which again causes the path starting at 
$g_3$ to change. What we are left with when we are finished are two constant 
paths, one at $g_1$ and one at $g_2$, along with a path which starts at $g_3$ 
and ends at $g_1g_2$. That is, the gauge field holonomy is constant (and equal 
to the appropriate boundary value) on each of the two upper legs of $P$, and 
only changes nontrivially on the lower leg. This final history, which
we denote  
by ${\tilde q_p}$, lies in the subspace $\phi (Q^H(C))$. Since our initial 
history $q_p$ was generic, we have shown that each path-component of $Q^H(P)$ 
contains {\it at least one} path-component of $\phi (Q^H(C))$. In order to 
demonstrate that $\pi_0(Q^H(P))=\pi_0(Q^H(C))$, it remains to be shown that 
each path-component of $Q^H(P)$ contains {\it only one} path-component of 
$\phi (Q^H(C))$. To see this, note that a small change in the initial history 
$q_p$ will lead to a correspondingly small change in the final 
history ${\tilde q_p}$. Therefore, moving $q_p$ around its path-component in 
$Q^H(P)$ simply moves ${\tilde q_p}$ around its corresponding path-component 
in $\phi (Q^H(C))$. Now suppose that a given component of $Q^H(P)$ contains
two components of $\phi (Q^H(C))$. If $q_p$ lies in either of these two 
components, the above deformation process will yield ${\tilde q_p}=q_p$ since 
$q_p$ already has constant holonomies on the two upper legs of $P$. But we can 
move $q_p$ continuously through $Q^H(P)$ from one component of $\phi (Q^H(C))$
to the other. This implies that ${\tilde q_p}$ also moves from one component of
$\phi (Q^H(C))$ to the other, contradicting the earlier result that continuous
changes in $q_p$ cannot change the the path-component of ${\tilde
q_p}$. Hence,  
each path-component of $Q^H(P)$ contains exactly one path-component of $\phi 
(Q^H(C))$, and the map $\phi_*:\pi_0(Q^H(C))\to\pi_0(Q^H(P))$ is a one-to-one 
correspondence. In other words, $\pi_0(Q^H(P))=\pi_1(G)$. 

\centinsert{
   \pscaption{
        \psannotate{\psboxto(0cm;5cm){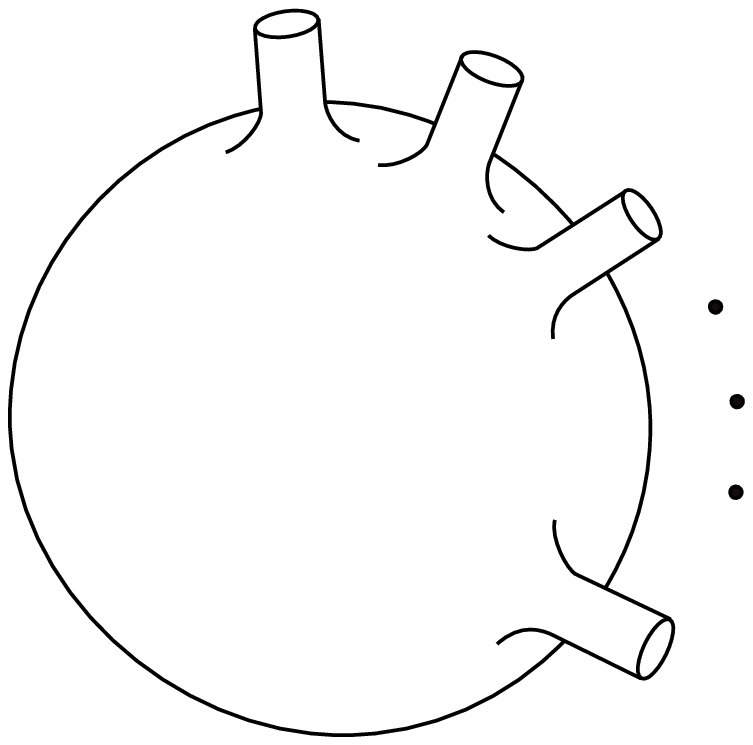}}
        {
         \at(11\pscm;9\pscm){1}
         \at(13\pscm;8.5\pscm){2}
         \at(15\pscm;6.4\pscm){3}
         \at(14.7\pscm;1.3\pscm){$n$}
        }}
{Fig. 2: The space-time $X_n$.}}

It is easy to demonstrate that this result still holds if we add additional
legs to the manifold $P$. More precisely, let $X_n$ denote a space-time 
with $n$ cylindrical legs and $n$ circular ends as shown in Fig.2. Both of 
the previous cases considered correspond to special values of $n$ --- 
namely, $X_2=C$ and $X_3=P$.  We can always choose a time-slicing of $X_n$ 
such that the associated basepoint network consists of a single junction 
with $n$ attached legs. We continue to require the elements of ${\cal G}^H$
to be trivial along this entire network. An element of $Q^H(X_n)$ can then 
be viewed as a set of $n$ paths in $G$ (representing the holonomies on each 
of the $n$ legs of $X_n$) starting at the points $g_1,g_2,\dots ,g_n$ 
(representing the $n$ boundary holonomies) and ending at some values $h_1,
h_2,\dots ,h_n$ (representing the $n$ holonomies at the junction). If we 
choose the $n$-th leg of $X_n$ as ``incoming'' and the rest as ``outgoing'', 
then there is also the junction constraint $h_n=h_1\cdots h_{n-1}$. 
As before, we can continuously deform any such history so that the holonomies 
are constant on every leg except (say) the incoming one. That is, at 
the end of the deformation we have one path in $G$ starting at $g_n$ and 
ending at the product $g_1g_2\cdots g_{n-1}$, and $n-1$ constant paths at 
the points $g_1,\dots ,g_{n-1}$. The same reasoning which we used to relate
the $n=2$ and $n=3$ cases then shows that the spaces $Q^H(X_n)$, $n\geq 2$, 
all have the same $\pi_0$. Note that one can obtain the space $X_{n-1}$ by 
simply ``closing off'' one of the boundaries in $X_n$. Similarly, if we 
choose any one of the boundary holonomies on $X_n$ to be the identity element 
of $G$, then the gauge theory on $X_n$ becomes equivalent to a gauge theory 
on $X_{n-1}$. In this way we see that $\pi_0(X_1)$ ($X_1$ is a disk) and 
$\pi_0(X_0)$ ($X_0$ is a sphere) are the same as $\pi_0(X_n)$, $n\geq 2$. The 
theories on $X_1$ and $X_0$ simply correspond to a special choice of boundary 
conditions in the higher $n$ cases. 

\centinsert{
   \pscaption{
        \psannotate{\psboxto(0cm;5cm){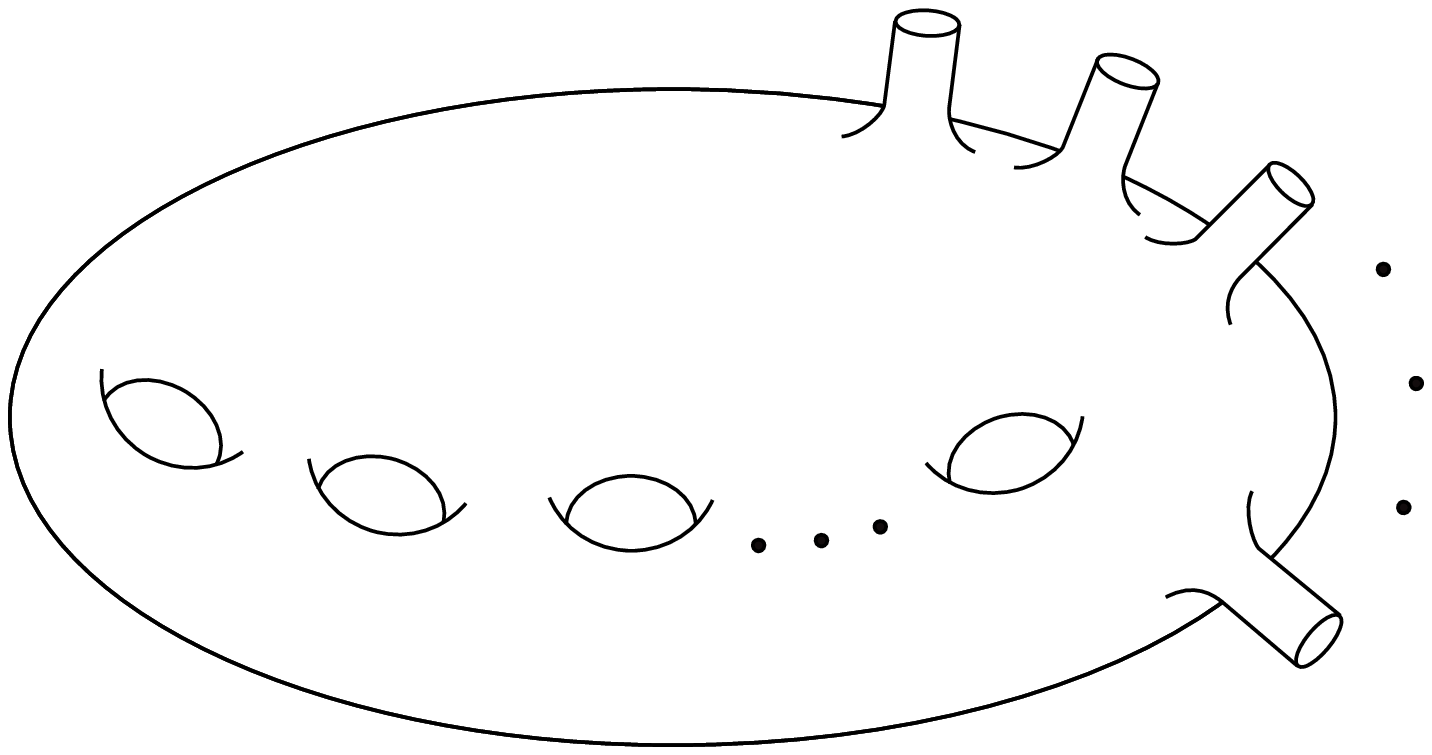}}
        {
         \at(5.0\pscm;2\pscm){1}
         \at(7.0\pscm;1.3\pscm){2}
         \at(10.0\pscm;1\pscm){3}
         \at(13.2\pscm;1.7\pscm){$r$}
         \at(12.0\pscm;8\pscm){1}
         \at(14.0\pscm;7.3\pscm){2}
         \at(15.5\pscm;6.5\pscm){3}
         \at(15.7\pscm;0.5\pscm){$n$}
        }}
   {Fig. 3: The space-time $X_{n,r}$.}}

\centinsert{
   \pscaption{
        \psannotate{\psboxto(12cm;0cm){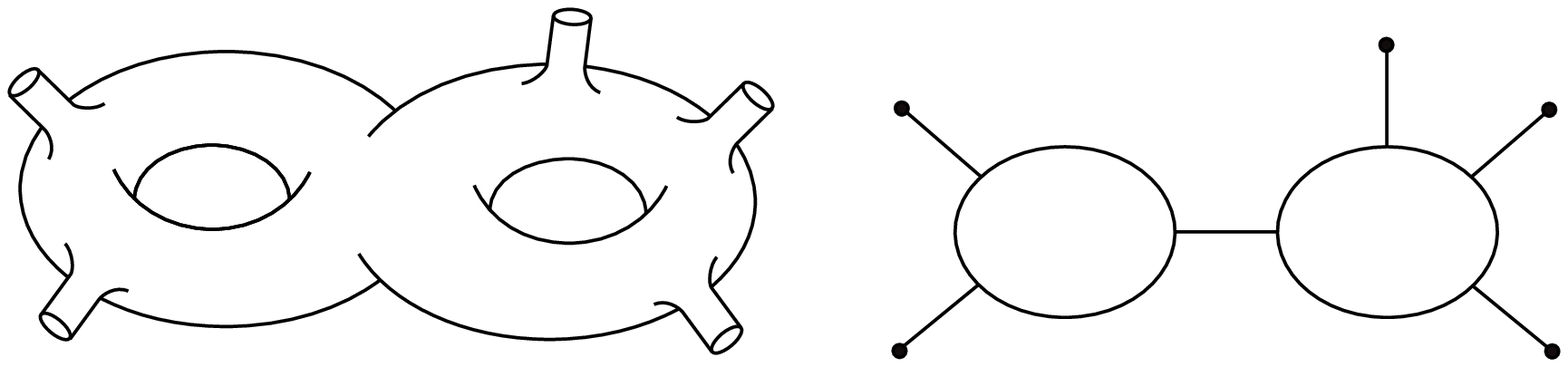}}
        {
         \at(5\pscm;0\pscm){(a)}
         \at(14\pscm;0\pscm){(b)}
        }}
   {Fig. 4: (a) shows the space-time $X_{5,2}$. The basepoint network 
associated with a particular time-slicing of $X_{5,2}$ is illustrated in (b).}}

More generally, we may add to $X_n$ an arbitrary number of ``handles'' as
in Fig.3. We denote this space-time with $n\geq 0$ boundaries and $r\geq
0$ handles (or {\it genus} $r$) by $X_{n,r}$. (Note that $X_{n,0}=X_n$.)
In any particular time-slicing, the space of histories $Q^H(X_{n,r} )$ 
will now consist of a certain class of maps from a complicated 
one-dimensional basepoint network into the gauge group $G$. This network 
will have $n$ external legs, $r$ internal ``loops'' and numerous
junctions. 
(As an example, a possible basepoint network for $X_{5,2}$ is shown in Fig.4.) 
As usual, there will be a constraint on the product of the holonomies at each 
junction, and we have required the elements of ${\cal G}^H$ to be
trivial along 
the entire network. The computation of $\pi_0(Q^H(X_{n,r}))$ is similar to the 
genus zero case above, but with one additional subtlety. We outline the 
procedure below. First, given any $q_c\in Q^H(C)$, we can still find a history 
in $Q^H(X_{n,r})$ which is equal to $q_c$ on one external cylindrical leg of 
$X_{n,r}$ and has constant holonomy on each other cylindrical portion (both 
internal and external). However, the values of these constants are no longer 
completely fixed by the combination of the boundary holonomies and the
junction 
constraints. There will be one unconstrained constant for each of the $r$ 
handles --- that is, one for each internal loop in the corresponding basepoint 
network. 
(These are similar to unconstrained loop momenta in a Feynman diagram.) 
If we want a continuous and one-to-one map $\phi :Q^H(C)\to
Q^H(X_{n,r})$ as we  
had in the $r=0$ case, then an arbitrary choice must be made for these $r$ 
``loop holonomies''. However, once these holonomies are fixed 
we can no longer give a
well-defined, continuous procedure for deforming an arbitrary history in
$Q^H(X_{n,r})$ into the subspace $\phi (Q^H(C))$, since each history
will have its own natural values for these constants.
But
there is a continuous and one-to-one map $f:Q^H(C)\times G^r\to Q^H(X_{n,r})$, 
where $G^r=G\times\cdots\times G$ ($r$~times), which avoids this problem. We 
simply include the unconstrained loop holonomies as extra degrees of
freedom in  
the domain of the map. That is, if we specify a cylindrical history in
$Q^H(C)$  
along with $r$ elements of $G$ (representing the loop holonomies), then $f$ 
gives us a history on
$X_{n,r}$ whose holonomy only changes on one 
external leg. In a manner similar to the $r=0$ case, we can now give a 
well-defined procedure for deforming any element of $Q^H(X_{n,r})$ 
into the subspace defined by the image of $f$, and use this to show that 
$\pi_0(Q^H(X_{n,r}))$ is in one-to-one correspondence with $\pi_0(Q^H(C)\times 
G^r)$. As $G$ is assumed connected ($\pi_0(G)=\{ e\}$), this yields
$\pi_0(X_{n,r})=\pi_0(Q^H(C))=\pi_1(G)$. Note that this result also applies to 
the closed surfaces $X_{0,r}$ since the gauge theory on such a surface can be 
obtained from the theory on $X_{n,r}$ (for any $n\geq 1$) by choosing the 
boundary holonomies to be trivial. Since any compact, orientable surface is 
homeomorphic to some $X_{n,r}$, we have shown: 
\vskip 12pt 
\noindent 
{\it For a two-dimensional pure Yang-Mills theory with (connected) gauge 
group $G$ defined on a compact, orientable space-time $M$, $\pi_0(Q^H(M))
=\pi_1(G)$ independent of the topology of $M$.}

\vskip 12pt 
\noindent
We can still 
view this correspondence as being induced by a map $\phi:Q^H(C)\to 
Q^H(X_{n,r})$ which is obtained from ${f:Q^H(C)\times G^r\to Q^H(X_{n,r})}$ 
by making some fixed choice of the $r$ loop holonomies. That is, 
${\phi_*:\pi_0(Q^H(C))\to\pi_0(Q^H(X_{n,r}))}$ is one-to-one and onto. Since 
$C$ may be viewed as the cylindrical leg of $X_{n,r}$ onto which we are 
deforming all the nontrivial holonomy changes, and any quantization of
the full  
gauge theory must yield an allowed quantization in this local region, we see 
that the inequivalent quantizations of the gauge theory on $X_{n,r}$ (for any 
$n$ and $r$) are labelled by the IUR's of $\pi_1(G)$. 

It is worth mentioning that although we have always used a particular 
time-slicing of $M$ in order to perform our analysis, the above result holds 
for {\it any} such choice. That is, $\pi_0(Q^H(M))$ is independent of the 
choice of time-slicing. Our result in the $r>0$ case {\it does}, however, 
depend critically on the assumption that $G$ is connected. For gauge groups 
with more than one path-component, such as $O(N)$, a more careful analysis 
yields $\pi_0(Q^H(X_{n,r}))=\pi_1(G)\times\pi_0(G)^{2r}$ (when the elements 
of ${\cal G}^H$ are required to be trivial only at a single point on 
$X_{n,r}$), and $\pi_0(Q^H(X_{n,r}))=\pi_1(G)\times\pi_0(G)^r$ (when the 
elements of ${\cal G}^H$ are required to be trivial along the entire 
basepoint network). Note that when $G$ is connected we obtain 
$\pi_0(Q^H(X_{n,r}))=\pi_1(G)$ for either of these two restrictions on the 
gauge transformations in ${\cal G}^H$. Therefore, as alluded to earlier, 
the results of this paper remain unchanged if we assume that space-time has 
only a {\it single} special basepoint rather than a network of such points. 
But our results {\it do} change if there is not at least one special point. 
More specifically, if $M$ is {\it a priori} compact then there is no natural 
place at which we can require the gauge transformations in ${\cal G}^H$ to 
be trivial. As discussed previously in the canonical case, we are thus 
forced to include {\it global} gauge transformations in ${\cal G}^H$. The 
space of histories $Q^H(M)$ will now be much more complicated. However, our 
analysis still applies to a certain subclass of quantizations of the theory 
which remain in one-to-one correspondence with the IUR's of $\pi_1(G)$ for 
all connected $G$ and compact, orientable space-times $M$. 

Those readers who are more familiar with homotopy theory may have realized 
that there are slicker methods for obtaining $\pi_0(Q^H(M))$ which completely 
avoid the need to perform a time-slicing. (For a general discussion of these 
methods, along with further references, see \ref\avis{S.~J.~Avis and 
C.~J.~Isham, in {\it Recent Developments in Gravitation}, edited by M.~Levy 
and S.~Deser (Plenum, New York, 1979)\semi R.~Percacci, {\it Geometry of 
Nonlinear Field Theories} (World Scientific, Singapore, 1986).}.) In 
particular, using such techniques it can be shown that (for connected $G$,
and $M$ possessing a special basepoint) 
$\pi_0(Q^H(M))=H^2({\tilde M};\pi_1(G))$. Here, ${\tilde M}$ is the surface 
obtained by closing up the boundary circles of the compact two-dimensional 
space-time $M$, and $H^2({\tilde M};\pi_1(G))$ is the second cohomology group 
of ${\tilde M}$ with coefficients in $\pi_1(G)$. The elements of 
$H^2({\tilde M};\pi_1(G))$ label the principal $G$-bundles over the closed 
surface ${\tilde M}$, and the gauge fields belonging to the path-component 
of $Q^H(M)$ associated with any such element can be viewed as connections on 
the corresponding bundle. If $M$ is orientable, then 
$H^2({\tilde M};\pi_1(G))=\pi_1(G)$ independent of the genus and the number 
of boundary components of $M$, and we recover the above results. Our main 
reason for performing the analysis using the more elementary time-slicing 
methods (and holonomies rather than gauge fields) is that we feel it makes the 
results more transparent by relating everything back to the cylindrical case. 
There is, however, something new that we can learn from the more advanced 
methods. More specifically, the relation $\pi_0(Q^H(M))=H^2({\tilde
M};\pi_1(G))$  
still holds even if $M$ is a {\it nonorientable} surface (such as the 
projective plane, the M\"obius strip, or the Klein bottle). In this case, 
$H^2({\tilde M};\pi_1(G))$ is isomorphic to the quotient group 
$\pi_1(G)/2\pi_1(G)$, where the normal subgroup $2\pi_1(G)$ consists of all 
squares of elements in $\pi_1(G)$. (Modding out by this normal subgroup is 
equivalent to adding relations to $\pi_1(G)$ which state that every element is 
equal to its inverse.) The inequivalent quantizations of these theories are 
then labelled by the IUR's of $\pi_1(G)/2\pi_1(G)$ independent of the details 
of the compact, nonorientable space-time $M$. This result can, in principle, 
also be obtained using the time-slicing methods. However, the reasoning is 
somewhat more subtle. 

We close with two comments. First, we are currently in the process of
addressing  
many of the questions which have already been studied in the $\theta =0$ case, 
only now for nonzero $\theta$. For example, can we still obtain the exact 
partition functions when $\theta\neq 0$? Are these theories still string 
theories for the appropriate gauge groups? Is there still a large-N phase 
transition as one moves from weak to strong coupling on the space-times $S^2$ 
and $P^2$? The answers to these questions may provide insights into 
four-dimensional gauge theories in nontrivial $\theta$-vacua. Second, we note 
that our study of $\theta$-vacua in 2D Yang-Mills theories on an arbitrary 
space-time is an application of a much more general procedure for relating 
topology change and inequivalent quantizations which will be outlined in a 
forthcoming paper \ref\tsi{T.~D.~Imbo and P.~Teotonio-Sobrinho, paper in 
preparation.} (see also related work in \ref\tc{A.~P.~Balachandran,
G.~Bimonte,  
G.~Marmo and A.~Simoni, Nucl. Phys. {\bf B446} (1995) 299.}). 

\bigskip

\centerline{{\bf Acknowledgements}}
\bigskip
\noindent
It is a pleasure to thank A.~P.~Balachandran, Lee Brekke, Sterrett
Collins, Alex King and Eric Martell for useful discussions. This research was 
supported in part by the U.~S.~Department of Energy under contract number 
DE-FG02-91ER40676. Most of this work was done while P.~T.-S. was a 
postdoctoral fellow at the University of Illinois at Chicago.

\listrefs
\bye